\documentclass[twocolumn,showpacs,prl,aps,superscriptaddress,amsmath,amssymb,floatfix]{revtex4}

\usepackage{graphicx}

\begin{document}

\title{Tunneling time scale of under-the-barrier forerunners}
\author{Gast\'on Garc\'{\i}a-Calder\'on}
\altaffiliation{{\bf Senior Associate ICTP}}
\email{gaston@fisica.unam.mx}
\affiliation{Instituto de F\'{\i}sica,
Universidad Nacional Aut\'onoma de M\'exico,
Apartado Postal {20 364}, 01000 M\'exico, Distrito Federal, M\'exico}
\author{Jorge Villavicencio}
\email{villavics@uabc.mx}
\affiliation{Facultad de Ciencias,
Universidad Aut\'onoma de Baja California,
Apartado Postal 1880, 22800 Ensenada, Baja California, M\'exico}

\date{\today}

\begin{abstract}
Time-dependent analytical solutions to Schr\"{o}dinger's equation with 
quantum shutter initial conditions  are used to investigate the issue of the
tunneling time of forerunners in rectangular potential barriers.
By using a time-frequency analysis, we find the existence of a regime characterized
by the opacity of the barrier, where the maximum peak of a forerunner  measured
at the barrier transmission edge $x=L$ corresponds to  a genuine tunneling process.
The corresponding time scale represents the tunneling time of the forerunner through
the classically forbidden region.
\end{abstract}

\pacs{03.65Bz,0365.Ca.}

\maketitle
Tunneling, that describes the possibility that a particle traverses through a classically forbidden region, is one of the paradigms of quantum mechanics. In the energy domain one solves Schr\"odinger's equation at a fixed energy $E$ to obtain the probability of transmission through a barrier region, a subject discussed in every quantum mechanics textbook. In the time domain, however, there are still aspects open to scrutiny. 
A problem that has remained controversial over the years is the tunneling time problem, that may be stated by the  question : How long does it take to a particle to traverse a classically forbidden region?  Different authors have proposed and defended different views in answering the above question\cite{traversal}. Some authors have argued that it might not be a unique definition of the tunneling time since different arrangements for the tunneling process may lead to different relevant time scales. An interesting idea that arises
from the above considerations is that the relevance of the different tunneling time definitions may depend on the context where they may become physically significant quantities\cite{gcvdm02}.

In recent work we have investigated the effect of the transient solutions to the time-dependent
Schr\"odinger's equation for cutoff wave initial conditions (quantum shutter) on the tunneling process\cite{gcr97,gcv01,gcv02}. In particular for tunneling through a barrier
we found that just across the tunneling barrier, the probability density as a function of time may exhibit a sharp peak maximum that we called {\it time domain resonance}. The peak value $t_{max}$ of that forerunner represents the largest probability of finding the particle at the barrier width $L$. We found also that the behavior of $t_{max}$ as a function of the barrier width $L$ exhibits  for small values of $L$ a  {\it basin} region,  followed by a region where $t_{max}$ grows linearly as $L$ increases\cite{gcv01}. More recently, in collaboration with Delgado and Muga\cite{gcvdm02}, we have investigated the time scale for forerunners  preceeding the main tunneling signal of a wave created by a source with a sharp onset\cite{mb00,vrs02} and by a quantum shutter\cite{gcv01} for systems with very large or infinite barrier width.
For the particular case of opaque finite barriers \cite{gcvdm02} we found a basin regime occurring at fixed  positions $0<x \ll \kappa_0^{-1}$ along the internal region, where forerunners are dominated by under-the-barrier frequency components. That is, in opaque barrier systems the tunneling forerunners are observed at distances of the order or smaller than the penetration length $\kappa_{0}^{-1}$, where $\kappa_{0}=[2m(V-E)]^{1/2}/\hbar$, with $E$ and $V$ corresponding to the incidence energy and the barrier height, respectively. However, at distances $\kappa_0^{-1} < x \leq L$ the non-tunneling components eventually dominate the time evolution process, and hence it is not possible to speak of a genuine tunneling time scale of forerunners at $x=L$. We believe that in order to gain more insight on the properties of  transient tunneling structures and their corresponding time scales, the existence of tunneling forerunners at the barrier edge $x=L$ needs to be investigated.

The aim of this work is to show the existence of a regime along the {\it basin} region,
where the maximum $t_{max}$ of the {\it time domain resonance} measured at the barrier width $L$ corresponds to under-the-barrier tunneling and consequently provides a genuine time scale for tunneling.

Our approach to the tunneling time problem is based on a model that 
deals with an explicit solution\cite{gcr97} of time-dependent
Schr\"{o}dinger equation for an arbitrary potential $V(x)$ ($0\leq x\leq L$)
that vanishes outside the internal region. We consider the problem 
of the time evolution of a cutoff plane wave $\Psi (x,k;t=0)=\Theta
(-x)(e^{ikx}-e^{-ikx})$ following the instantaneous opening at $t=0$ of
a quantum shutter at $x=0$. 
Along the tunneling region the solution reads,
\begin{eqnarray}
\Psi^i(x,k,t)=&&\Phi_k(x)M(y_k)-\Phi_{-k}(x)M(y_{-k})\nonumber\\
&&-\sum_{n=-\infty}^{\infty} \Phi_n(x) M(y_{k_{n}}), \,\,\,(0 \leq x \leq L)\,\,\,\,\,
\label{psii}
\end{eqnarray}
where the $\Phi_{\pm k}(x)$'s refer to the stationary solutions of the problem and 
$\Phi_n(x)=2iku_n(0)u_n(x)/(k^2-k_n^2)$.
Similarly, the solution $\Psi^e(x,k;t)$ for the external or transmitted region 
($x \geq L$), is given by \cite{gcv01},

\begin{eqnarray}
\Psi^e(x,k;t)=&&T_{k}M(y_k)-T_{-k}M(y_{-k})\nonumber\\
&& -i\sum\limits_{n=-\infty }^{\infty }T_{n}M(y_{k_{n}}), \,\,\,\,\,(x \geq L) 
\label{psie}
\end{eqnarray}
where the $T_{\pm k}$'s refer to the transmission amplitudes, and the
factor $T_{n}=2iku_{n}(0)u_{n}(L)\exp (-ik_{n}L)/(k^{2}-k_{n}^{2})$. In Eqs.\ 
(\ref{psii}) and (\ref{psie}) the coefficients $\Phi_n$ and $T_n$ are given
in terms of the resonant eigenfunctions $\{u_{n}(x)\}$ with complex
energy eigenvalues $E_n=\hbar^2k_n^2/2m$, with $k_{n}=a_{n}-ib_{n}$
($a_{n},b_{n}>0$). The resonant sums in Eqs.\ (\ref{psii}) and(\ref{psie})
run over the full set of complex poles $\{k_{n}\}$. The $M^{\prime }s$ are defined as \cite{gcr97},
\begin{equation}
M(y_{q})=\frac{1}{2}e^{imx^{2}/2\hbar t}w(iy_{q})
\label{m}
\end{equation}
where $w$ is the complex error function \cite{abrwtz}defined as 
$w(z)=\exp (-z^{2})erfc(-iz)$, with arguments 
$y_{q}(x,t)=e^{-i\pi /4}(m/2\hbar t)^{1/2}[x-\hbar qt/m]$, 
where $q=\pm k,$ $k_{\pm n}$.

\begin{figure}[!tbp]
\rotatebox{0}{\includegraphics[width=3.3in]{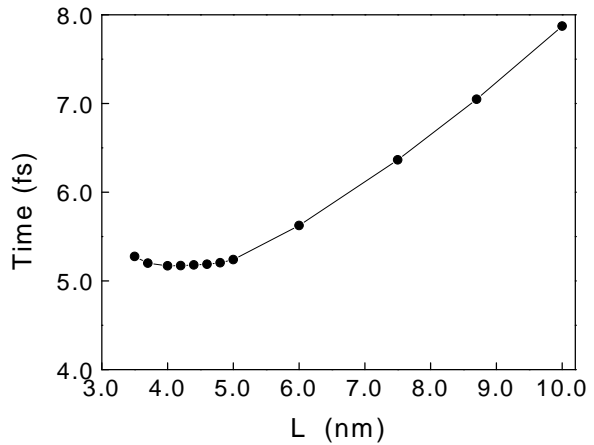}}
\caption{ Maximum of the {\it time domain resonance} $t_{max}$ 
(full dot) as a function of the barrier width $L$, for an 
incidence energy $E=.001$ eV. In this case the barrier height 
is $V=0.3$ eV. See text.}
\label{baslin}
\end{figure}

We shall explore the issue of the tunneling time  scale
in typical one dimensional potential barriers \cite{mendez} of height
$V$ and thickness $L$, defined along the interval $0\leq x\leq L$. 
Let us analyze the behavior of the maximum of the 
{\it time domain resonance}, $t_{max}$, as a function of 
the potential barrier width, $L$. Let us recall that this transient structure
corresponds to the first maximum of the probability density $|\Psi^e|^2$, 
measured at the fixed position, $x=L$. This can be appreciated in 
Fig. \ref {baslin} where we plot the $L-$dependence of $t_{max}$ 
(full dot), corresponding to potential barrier systems with
parameters: $V=0.3$ eV, incidence energy 
$E=\hbar^2k^2/2m=0.001$ eV, and effective mass for the electron 
$m=0.067m_e$. 
Here we can clearly observe a {\it basin} corresponding to a range of 
values of the barrier width. We can also 
appreciate that if $L$ is further increased, $t_p$ grows linearly with $L$. 
Such a linear regime occurs at large barrier 
widths. We have argued \cite {gcv01} that in this case, the tunneling process
is inhibited and that the particle goes mainly over the barrier.  
\begin{figure}
\rotatebox{0}{\includegraphics[width=3.3in]{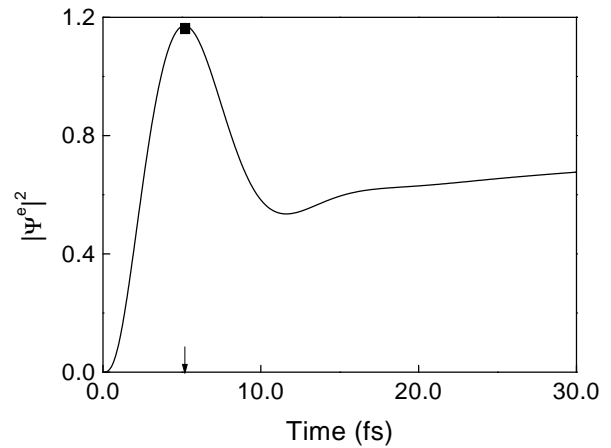}}
\caption{Time evolution of $|\Psi^e|^2$ (solid line), at $x=L=4.0$ nm.
A full square indicates the position of the maximum of the 
{\it time-domain resonance}, at $t_{max}=5.17$ fs.}
\label{tdr}
\end{figure}
We have recently suggested \cite{gcv01} that for small values of 
the barrier width $L$, the  {\it basin} exhibited 
by $t_{max}$ , is a result of a subtle interplay between tunneling
and top-barrier resonant processes. 
In what follows we shall investigate under what conditions the 
time scales associated to the {\it basin}, are in fact related to a tunneling 
process. We begin our analysis by choosing the case $L=4.0$ nm depicted in 
Fig. \ref{baslin}, which is located around the minimum of the {\it basin}. 
In Fig. \ref{tdr} we plot the normalized probability density $|\Psi^e|^2$ 
(solid line), normalized to the transmission coefficient $|T_k|^2$, as a 
function of time $t$ for a fixed value of position $x=L$. We can clearly 
appreciate a {\it time domain resonance} peaked at the value $t_{max}=5.17$ fs. 
\begin{figure}[!tbp]
\rotatebox{0}{\includegraphics[width=3.3in]{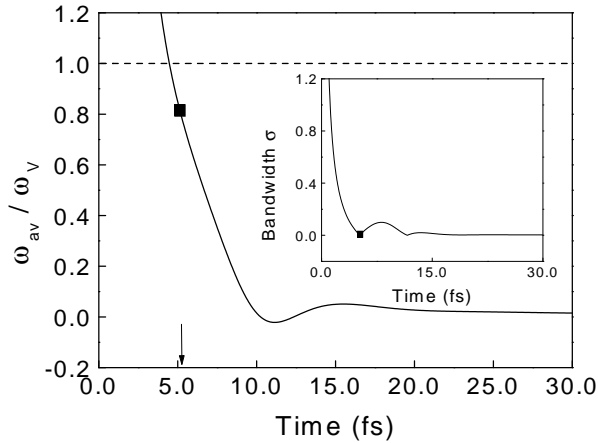}}
\caption{ Relative average local frequency $\omega_{av}/\omega_V$
(solid line) for the case depicted in Fig. \ref{tdr}. 
The cutoff-frequency $\omega_V/\omega_V=1$ (dashed line) is included 
for comparison. 
In the inset we plot the instantaneous bandwidth $\sigma$ 
of the spectrogram depicted in the main graph. Notice that the
frequency deviations from $t_{max}$ are exactly zero, {\it i.e.},
$\sigma(t_{max})=0$. In all cases a full square indicates the 
position of $t_{max}$.}
\label{espectro}
\end{figure}
In order to determine if this structure is related to a genuine
tunneling process we analyze the frequency content of $|\Psi^e|^2$. 
We compute the local average frequency $\omega_{av}$ \cite{mb00,cohen}, 
\begin{equation}
\omega_{av}=-{\rm Im}\left[\frac{1}{\Psi^s} \frac{d}{dt}\Psi^s
 \right],
\label{4}
\end{equation}
and the instantaneous bandwidth $\sigma$\cite{cohen},

\begin{equation}
\sigma=\left |{\rm Re} \left  [\frac{1}{\Psi^s}\frac{d}{dt}\Psi^s \right ]\right |,
\label{5}
\end{equation}
where $s=i,e$ according to consider, respectively, the internal or external solutions.

In Fig. \ref{espectro} we plot the relative average local frequency 
(relative frequency for short) $\omega_{av}/\omega_V$, where 
$\omega_V=V/\hbar$ is the cut-off frequency,  along the relevant time 
interval, discussed in Fig. \ref{tdr}. We can appreciate that in the 
vicinity of the maximum of the {\it time-domain resonance}, $t_{max}$, 
the probability density is composed  entirely by under-the-barrier
frequency components {\it i.e}  $\omega_{av}/\omega_V<1$. This also occurs 
at the exact value $t_{max}$, also indicated in the figure by a solid square. 
In the inset of Fig. \ref{espectro} we plot the instantaneous bandwidth $\sigma$ of the spectrogram. Notice the absence of a frequency dispersion around the maximum $t_{max}$, {\it i.e.}, $\sigma(t_{max})=0$. The above result indicates that in this case the peak of the {\it time domain resonance}, and the values close to it, refer to a genuine tunneling event.
\begin{figure}[!tbp]
\rotatebox{0}{\includegraphics[width=3.3in]{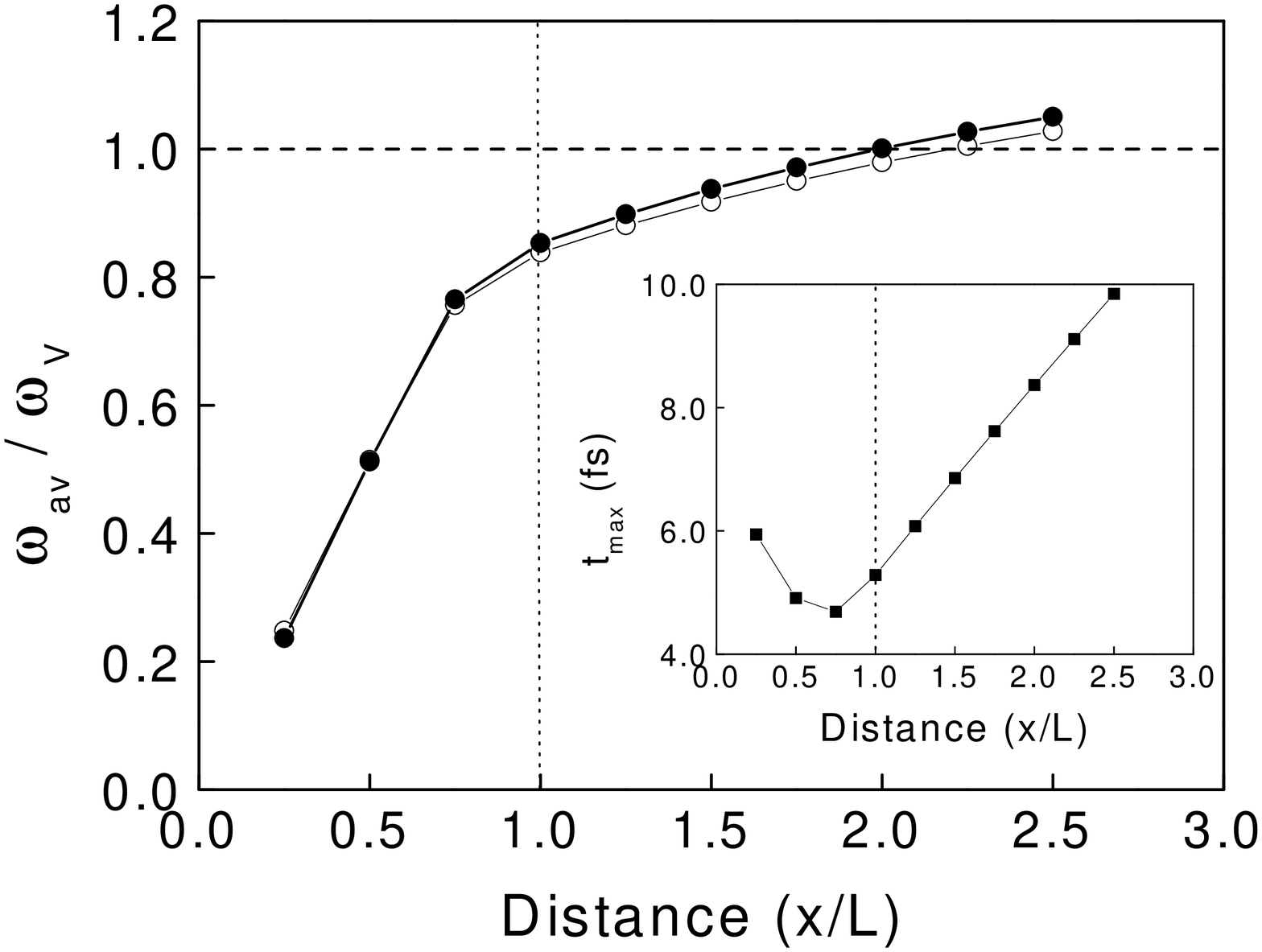}}
\caption{ Relative frequency $\omega_{av}/\omega_V$
of the maximum of the {\it time-domain resonance}, 
as a function of position, measured in units of the barrier width
$L$. The parameters are given in the text. Two incidence energies are considered: 
$E=0.001$ eV (solid dot), and $E=0.01$ eV (hollow dot). In both 
cases, the relative frequency $\omega_{av}/\omega_V$ along the 
internal region is below the cutoff-frequency $\omega_V/\omega_V=1$ 
(dashed line). The behavior of $t_{max}$ as a function of position
is illustrated in the inset for the case with $E=0.01$ eV.
The position of the barrier edge, $x=L$, is indicated by 
a dotted line in both figures.}
\label{omgvsx}
\end{figure}
\begin{figure}[!tbp]
\rotatebox{0}{\includegraphics[width=3.3in]{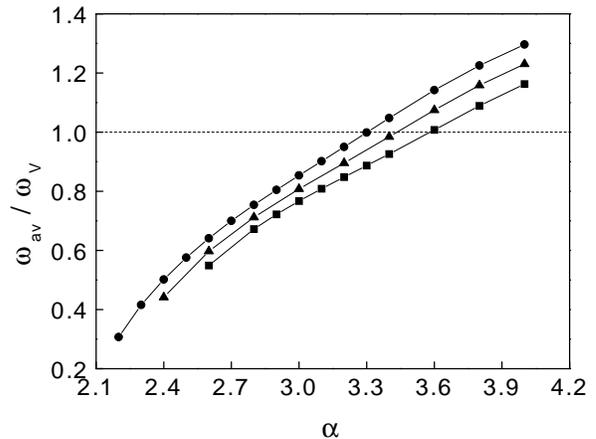}}
\caption{Relative frequency $\omega_{av}/\omega_V$ measured
at the barrier edge $x=L$, as a function of the opacity
$\alpha$. Here we considered a barrier height 
$V=0.3$ eV, and the parameters: $u=300$ (solid dot), $u=10$ 
(solid triangle), and $u=5$ (solid square). Note that for
values of the opacity smaller than $\alpha \simeq 3.3$, the relative
frequencies for all values of $u$ are below the cutoff-frequency
$\omega_V/\omega_V=1$(dashed line). See text.}
\label{omgvsalf}
\end{figure}
In Fig. \ref{omgvsx} we plot the relative frequency $\omega_{av}/\omega_V$
associated to different values of the maximum $t_{max}$
of the {\it time-domain resonance}, measured at different positions along
both the internal and external regions of a potential barrier with
parameters: $V=0.3$ eV, $L=4.13$ nm, and $\alpha=3.0$. 
In this case we choose the following values of the incidence energy: $E=0.001$ eV 
(solid dot), and $E=0.01$ eV (hollow dot). 
In the inset of Fig. \ref{omgvsx} we show, for the particular case of $E=0.01$ eV, 
the values of $t_{max}$ (solid square) at the different values of position considered
in the main graph.  
As can be clearly appreciated in that figure, the tunneling process along the whole internal region is governed by under-barrier-frequency components, {\it i.e.}, $\omega_{av}/\omega_V<1$. 
The above results should be contrasted with those obtained for very large or 
infinite barrier widths recently studied in Ref. \cite{gcvdm02}, where
under-the-barrier tunneling components arise only at distances smaller than 
a characteristic length. Here we have found a particular
combination of potential  parameters characterizing a genuine tunneling process.
The existence of under-the-barrier tunneling forerunners is not 
only restricted to values of the position along the internal region 
of the potential. In fact, we can see in Fig.\ref{omgvsx}, that 
for distances up to $x \simeq 2L$ along the external region, we can 
still observe frequency components below the cutoff-frequency, $\omega_V$.
As $ x/L$ increases further, $ \omega_{av}/\omega_V > 1$.  This behavior follows  
because the corresponding values of $t_{max}$ increase linearly, indicating the 
prevalence of non-tunneling components (see the inset in Fig. \ref{omgvsx}). 

We have found that the existence of tunneling forerunners 
at the barrier edge $x=L$ may be described more generally by 
referring to the opacity $\alpha$ of the system, defined as,
\begin{equation}
\alpha = \frac{\left [2mV\right]^{1/2}}{\hbar}L,
\label{op}
\end{equation}
and by the dimensionless parameter $u$, the ratio between the potential barrier 
height and the incidence energy, 
\begin{equation}
u=\frac{V}{E}.
\label{u}
\end{equation}

There is a regime, that we shall refer to as {\it tunneling regime},
characterized by the opacity $\alpha$ and  the parameter $u$, 
where the relative frequencies associated to the {\it time domain resonance}
are below the cutoff-frequency $\omega_V$. In order to show this, in Fig. 
\ref {omgvsalf} we plot the relative frequency $\omega_{av}/\omega_V$ as 
function of the opacity $\alpha$, for three different values of the 
parameter $u$, $u=5$ (solid dot), $u=10$ (solid triangle), 
and $u=300$ (solid square).  Although in this case we have chosen a value of 
$V=0.3$ eV, we have checked that the relative frequency $\omega_{av}/\omega_V$, exhibits 
a striking regularity when plotted as a function of $\alpha$. For a given value of 
$\alpha$ all the systems with the same parameter 
$u$, share the same curve. That is, all systems characterized by the same 
parameters $\alpha$ and $u$ yield the same relative frequency,  
$\omega_{av}/\omega_V$. In view of this observed regularity, we can 
completely characterize the  regime associated with under-the-barrier 
frequency components. Although in Fig. \ref {omgvsalf} we have considered 
values of the parameter $u$ such that $5 \leq u \leq 300$, the case 
corresponding to very large values of $u$ ($u\rightarrow \infty$), almost 
overlaps with the case $u=300$. Thus, for very large values of $u$ there is  
an upper-bound for the opacity $\alpha_u\simeq 3.3$. This result,
and the fact that the lower-bound for the opacity \cite {notelb} is given by 
$\alpha_c=2.065$, recently reported in Ref. \cite{gcv02}, allows to characterize an 
opacity ``window" given by $\alpha_c \leq \alpha \leq \alpha_u$ where the relative
frequencies are always below the cutoff-frequency $\omega_V/\omega_V=1$, 
irrespective of the value of the parameter $u$, namely of the value of the incidence 
energy. Note that the values of $\alpha$ within the opacity ``window'' may be obtained using typical parameters of semiconductor heterostructures\cite{mendez}. From the figure one can appreciate 
the existence of another regime of $\alpha$'s where the existence of under-the-barrier processes associated to the {\it time-domain resonance} may depend on the value 
of the parameter $u$. 

It is of interest to remark that the time scale given by the peak maximum $t_{max}$ of the forerunner at the barrier edge $x=L$, i.e., the {\it time domain resonance}, differs from both the semi-classical B\"uttiker-Landauer and B\"uttiker traversal times, which exhibit a linear dependence with $L$, and also differs from the phase-time\cite{gcv01,gcv02}. In Ref. \cite{gcv02} we have discussed an unexpected relationship between the delay time and the existence of {\it time domain resonances}.

To conclude we  remark that the analytical solution to the time-dependent Schr\"odinger's equation with quantum shutter initial conditions applies in general to arbitrary potentials provided they vanish beyond a distance and can also be extended to deal with cutoff pulses
as discussed in Ref. \cite{gcv02}. Our results refer to novel transient effects in time-dependent tunneling. To test our results experimentally would require to consider the detection of tunneling particles in time domain at distances close to the interaction region. 

\acknowledgments{The authors thank J. G. Muga for a critical reading of the manuscript and acknowledge financial support of DGAPA-UNAM under grant No. IN101301.}

\end{document}